\documentclass[twoside,12pt]{article}
\usepackage{amsmath}
\usepackage{amssymb}
\usepackage{latexsym}
\usepackage{epsfig}
\usepackage{cite}
\newlength{\dinwidth}
\newlength{\dinmargin}
\newtheorem{theorem}{Theorem}[section]
\newtheorem{prop}[theorem]{Proposition}
\newtheorem{lemma}[theorem]{Lemma}

\newtheorem{definition}{Definition}
 
\newenvironment{proof}{\medskip \noindent 
            {\bf Proof.}}{ \hfill $\square$ \medskip}

\newcommand{\wm}{{W\!\!\!\!\text{\raisebox{1.8ex}[0ex][0ex]%
{$\scriptscriptstyle\sim$}}}}
\def\wMin{{\Ws\!\!\!\!\text{\raisebox{1.7ex}[0ex][0ex]%
{$\scriptscriptstyle\sim$\,}}}}

\def\WRM{{\wm}\,^{(1)}} 
\def\EM{{E\!\!\!\text{\raisebox{1.8ex}[0ex][0ex]%
{$\scriptscriptstyle\sim$\,}}}}
\def\ERM{\EM^{(1)}} 

\newcommand{\wrw}{{W\!\!\!\!\!\text{\raisebox{-0.7ex}[0ex][0ex]%
{$\scriptscriptstyle\sim$\,}}}}

\def\dcrw{{\Os\!\!\!\!\text{\raisebox{-0.7ex}[0ex][0ex]%
{$\scriptscriptstyle\sim$\,}}}} 
\newcommand{\dcrwsub}[1]{\dcrw{}_{#1}}   
\def\wRW{{\Ws\!\!\!\!\!\text{\raisebox{-0.7ex}[0ex][0ex]%
{$\scriptscriptstyle\sim$ }}}} 
\def\Arw{{\As\!\!\!\!\text{\raisebox{-0.8ex}[0ex][0ex]%
{$\scriptscriptstyle\sim$\,}}}} 
\def\Rrw{{\Rs\!\!\!\!\text{\raisebox{-0.7ex}[0ex][0ex]%
{$\scriptscriptstyle\sim$\,}}}}  
\def\wnetrw{\{\Arw(\wrw)\}_{\wrw \in \wRW}}
\def\wrnetrw{{\{\Rrw(\wrw)\}_{\wrw \in \wRW}}}
\def\orw{{\omega\!\!\!\!\text{\raisebox{-0.7ex}[0ex][0ex]%
{$\scriptscriptstyle\sim$\,}}}}
\def\Orw{{\Omega\!\!\!\!\text{\raisebox{-0.7ex}[0ex][0ex]%
{$\scriptscriptstyle\sim$\,}}}}

\def\wdS{{\Ws}}

\def\wrnetds{{\{\Rs(W)\}_{W \in\wdS}}}

\def\As{{\mathcal A}}
\def\Bs{{\mathcal B}}

\def\Hs{{\mathcal H}}

\def\Os{{\mathcal O}}

\def\Rs{{\mathcal R}}

\def\Ts{{\mathcal T}}

\def\Ws{{\mathcal W}}

\def\Lid{SO_0(4,1)}

\def\idty{{\leavevmode\hbox{\rm 1\kern -.3em I}}}

\def\RR{\mathbb{R}} 
\def\CC{\mathbb{C}} 

\newcommand{\Min}{$\RR^5$} 
\newcommand{\Minm}{\RR^5}
\newcommand{\RW}{\mbox{\it RW}\ }
\newcommand{\RWm}{\mbox{\it RW}}
\newcommand{\dS}{\mbox{\it dS}\ }
\newcommand{\dSm}{\mbox{\it dS}}

\newcommand{\pihalf}{{\frac{\pi}{2}}}

\newcommand{\eps}{{\varepsilon}}

\newcommand{\proprw}[1]{\alpha\!\!\!\!\text{\raisebox{-0.7ex}[0ex][0ex]%
{$\scriptscriptstyle\sim$\,}}_{#1}}   
\newcommand{\curw}[1]{x_{#1}}         
\newcommand{\dotcurw}[1]{\dot{x}_{#1}}
\newcommand{\cumin}[1]{\tilde{x}_{#1}}
\newcommand{\dotcumin}[1]{\dot{\tilde{x}}_{#1}}
\newcommand{\cuSO}[1]{\lambda_{#1}}   
\newcommand{\fadcrw}[1]{\dcrw({#1})}  
\newcommand{\Ad}{{\rm Ad }}
\newcommand{\id}{{\rm id }}
\newcommand{\gm}{{g\!\!\!\text{\raisebox{1.2ex}[0ex][0ex]%
{$\scriptscriptstyle\sim$}}}}         
\newcommand{\nablam}{{\nabla\!\!\!\!\text{\raisebox{1.9ex}[0ex][0ex]%
{$\scriptscriptstyle\sim$}}}}         
\newcommand{\am}{{a\!\!\!\text{\raisebox{1.2ex}[0ex][0ex]%
{$\scriptscriptstyle\sim$}}}}         
\newcommand{\barcurw}[1]{\bar{\,\curw{#1}}}   
\newcommand{\barcuSO}[1]{\!\bar{\;\cuSO{#1}}} 
\newcommand{\xm}{{\tilde{x}}}

\begin{document}

\title{Covariant and Quasi-Covariant Quantum Dynamics in 
Robertson--Walker Space--Times} 

\author{ Detlev Buchholz, Jens Mund   \\
Institut f\"ur Theoretische Physik, Universit\"at G\"ottingen,\\
Bunsenstra\ss e 9, D-37073 G\"ottingen, Germany\\
\vphantom{X}\\
{Stephen J.\ Summers } \\
Department of Mathematics, University of Florida,\\
Gainesville FL 32611, USA\\}
\date{}

\maketitle 

\begin{abstract}We propose a canonical description of the dynamics of quantum
systems on a class of Robertson--Walker space--times. We show that the
worldline of an observer in such space--times determines a unique orbit
in the local conformal group $SO(4,1)$ of the space--time and that this
orbit determines a unique transport on the space--time. For a quantum
system on the space--time modeled by a net of local algebras, the
associated dynamics is expressed via a suitable family of ``propagators''.
In the best of situations, this dynamics is covariant, but more typically
the dynamics will be ``quasi-covariant'' in a sense we make precise.
We then show by using our technique of ``transplanting'' states and
nets of local algebras from de Sitter space to Robertson--Walker space 
that there exist quantum systems on Robertson--Walker spaces with 
quasi-covariant dynamics. The transplanted state is locally passive, 
in an appropriate sense, with respect to this dynamics. 
\end{abstract}

\section{Introduction}

     We address the question of how the dynamics of quantum systems on
curved space--times can be described in a physically motivated and
mathematically useful manner in the special case of a class of
Robertson--Walker space--times. In the literature known to us, the
general question is broached either for free fields or in some
quasi-classical approximation. We would like to give a purely quantum
field-theoretic treatment which is valid also for interacting quantum
fields. We do so in the language of algebraic quantum field
theory\cite{Haag,Wald}, which, though not widely known, has the advantages
of being conceptually broad enough to subsume within it most
approaches to quantum field theory and yet being mathematically
rigorous.

     Corresponding to the operationally motivated nature of
algebraic quantum field theory, we begin our considerations with the
worldline of an observer. This moving observer is subject to various 
forces --- if his 
worldline is a geodesic, these forces are purely gravitational;
otherwise, he is subject to acceleration due to non-gravitational
causes. The same is true if the worldline is that of a quantum system,
subject now to forces on the quantum level.\footnote{We shall, however,
neglect the extreme situations where the back reaction of the quantum
system on the space--time must be taken into account. Thus, for our
purposes, it will suffice to consider quantum field theory on a
fixed classical space--time background.} What would be a suitable,
mathematically rigorous description of the attendant dynamics of this
system? 

     We propose an answer to that question, at least for quantum
systems on a large class of Robertson--Walker space--times, which
includes de Sitter space. In Section 2, we shall specify the class of
Robertson--Walker space--times we shall be considering. For all of
these space--times, the local conformal group is $SO(4,1)$. We show
that given an arbitrary worldline in such a space--time, two physically
motivated assumptions allow us to find a unique curve in $SO(4,1)$
which generates the worldlines of the given system and neighboring
systems. These considerations are purely geometrical in nature.

     Quantum theory enters for the first time in Section 3, where we
describe quantum systems on our class of Robertson--Walker
space--times using nets of local observable algebras supplied with a
state.  We define there what we mean by covariant, respectively
quasi-covariant, dynamics of such systems. This will involve
continuous 2-parameter families of automorphisms of the global
observable algebra satisfying canonical propagator identities,
which implement the action of the curve in $SO(4,1)$ associated with
the evolution of the initial worldline, and which act locally on the
observables in a well motivated manner.

     In Section 4, we shall use results from a previous paper
\cite{BMS} to construct nets and states on the specified
Robertson--Walker space--times which admit dynamics in the sense of
Section 3. We provide examples which admit covariant dynamics, as well
as examples which only admit quasi-covariant dynamics. The transplanted
state will be shown to be passive in an appropriate sense for this dynamics,
so that for this dynamics the state behaves like a local equilibrium state
from which cyclic processes cannot extract energy.

     In the final section, we shall discuss the relation between our
work and an alternate proposal for dynamics of quantum systems on 
curved space--times made by Keyl\cite{Keyl1,Keyl2}. We shall also
make some remarks about the further outlook of this program.
An appendix contains proofs of a more technical nature.

\section{Kinematics} 

     To begin, we specify the class of Robertson--Walker space--times
we shall be considering, simultaneously establishing notation we
shall be using throughout this paper. Robertson--Walker space--times
are Lorentzian warped products of a connected open subset $I$ of $\RR$
with a Riemannian manifold of constant sectional curvature \cite{BEE,HE,O}.
We restrict our attention here to the subclass of those having
positive curvature, which may be assumed to be equal to $+1$. The 
corresponding Robertson--Walker space--times are homeomorphic to 
$I \times S^3$, and one can choose coordinates so that the metric 
has the form 
\begin{equation}
ds^2 = dt^2 - S^2(t)d\sigma^2\, .
\end{equation}
Here, $t$ denotes time, $S(t)$ is a strictly positive smooth function and
$d\sigma^2$ is the time-independent metric on the unit sphere $S^3$:
\begin{equation}
d\sigma^2 = d\chi^2 + \sin^2(\chi)(d\theta^2 + \sin^2(\theta)d\phi^2)\, .
\end{equation}

     Following \cite{HE} we define a rescaled time parameter $\tau(t)$ by
\begin{equation} \label{eqtau}
d\tau / dt = 1/S(t)\, .
\end{equation}
In terms of this new variable, the metric takes the form 
\begin{equation}
ds^2 = S^2(\tau)(d\tau^2 - d\sigma^2)\, , \label{rwmetric}
\end{equation}
where $S(\tau)$ is shorthand for $S(t(\tau))$. Since $S$ is everywhere
positive, $\tau$ is a continuous, strictly increasing function of $t$;
its range is therefore an open interval in $\RR$. In this paper we
shall further restrict our attention to those Robertson--Walker 
space--times with functions $S(t)$ such that the range of values of 
$\tau$ is of the form $(-\alpha ,\alpha)$, with $\alpha \leq \pi/2$. 
Henceforth, we shall denote by \RW any one of this class of Robertson--Walker
space--times. 

     As is well known, the four--dimensional de Sitter space--time \dS
can be embedded into five--dimensional Minkowski space \Min{} as 
follows:
\begin{equation}
dS = \{ x \in \Minm \, \mid \, x_0^2 - x_1^2 - \ldots - x_4^2 = -1 \}\, .
\label{eqdSMin}
\end{equation}
This space--time is topologically equivalent to $\RR \times S^3$, and in
the natural coordinates the metric has the form
\begin{equation}
ds^2 = dt^2 - \cosh^2(t)d\sigma^2 \, .
\end{equation}
We recognize \dS as a special case of Robertson--Walker space--time with
the choice $S(t) = \cosh(t)$. Once again, we change time variables by
defining 
\begin{equation}
\tau = \arcsin(\tanh(t))
\end{equation}
(so that $d\tau /dt = 1/\cosh(t)$), which takes values
$-\frac{\pi}{2} < \tau < \frac{\pi}{2}$. Then the metric becomes
\begin{equation}
ds^2 = \cos(\tau)^{-2}(d\tau^2 - d\sigma^2)\, . \label{dsmetric}
\end{equation}
The isometry group, as well as the conformal group, of \dS is the de 
Sitter group $O(4,1)$.    

     Comparing equations (\ref{rwmetric}) and (\ref{dsmetric}), it is now
clear (and well known \cite{HE}) that each of the Robertson--Walker
space--times specified above can be conformally embedded into de Sitter
space--time, {\it i.e.} there exists a global conformal diffeomorphism
$\varphi$ (see Definition 9.16 in \cite{BEE}) from \RW onto a subset of
\dS. How large the embedding in \dS is depends on the range of
the variable $\tau$ in each case examined, which itself depends upon
the function $S(t)$. If $\alpha = \pi/2$, then the embedded \RW coincides
with \dS. In this case, the de Sitter group $O(4,1)$ is the conformal
group of \RW. If $\alpha < \pi/2$, then $O(4,1)$ is only locally the
conformal group of \RW, {\it i.e.} there is an action induced upon
\RW through the embedding, but some orbits of points under this action
reach infinity in finite ``time''. Henceforth, for the sake of brevity
of expression, we shall identify \RW with its embedding in \dS whenever
it is convenient.

     We can now turn to the first step of our program: given a
worldline\footnote{By a worldline we mean an inextendible timelike
curve of class $C^2$, not necessarily parametrized by proper time.}
of an observer, construct a curve in $SO(4,1)$ which generates a flow
describing the worldlines of the observer and his neighbors. This
construction will be based upon physically natural hypotheses, and the
resultant curve will turn out to be unique.

     Let $t\mapsto \curw{t}$ be the worldline in \RW of the observer, 
parametrized by proper time $t$. Let then $t\mapsto \cuSO{t}$ be a 
differentiable curve in the group $SO(4,1)$ such that 
\begin{equation} \label{eqgtx0}
   \cuSO{t}\,\curw{0}=\curw{t} \quad  \text{ for all } t\in \RR\,. 
\end{equation}
Such a curve always exists, but it is only fixed up to elements of the
stability group of $\curw{0}$. In order to remove this ambiguity, we 
bring to bear physical considerations. We require that the curve generates, 
in the same sense as in equation~\eqref{eqgtx0}, also the worldlines of 
{\it neighboring} material particles\footnote{$t$ is in general not the 
proper time for these particles.} moving along with the observer without 
colliding with him or rotating with respect to him. These are interpreted 
as material points in the observer's laboratory, which itself is assumed to 
be at rest relative to gyroscope axes carried by the observer. 
These considerations find their formal expression in the following
conditions:

\begin{enumerate}
\item[(a)] For each $y_0$ in some neighborhood of $\curw{0}$, the
  events $\cuSO{t}\,y_0$, $t\in \RR$, describe,  potentially, 
the worldline of some material particle. This worldline is either
disjoint from the observer's worldline or coincides with it. 
\item[(b)] For given $y_0$ spacelike to $x_0$, the axis of a gyroscope 
carried by the observer at the space-time
point $\cuSO{t}\,x_0$ points towards the point $\cuSO{t}\,y_0$ at all times
$t$. 
\end{enumerate}

\noindent It turns out that these mild assumptions, in addition to  
equation~\eqref{eqgtx0}, uniquely fix $t\mapsto \cuSO{t}$. We
sketch the elements of the argument and state the theorem here;
the proof can be found in the appendix.

     Since only the conformal structure is involved, we may 
equally well consider the (conformally equivalent) \dS metric.  
Our assumptions imply that the curve  $t\mapsto \cuSO{t}$ generates a local
flow, which is irrotational and rigid with respect to the \dS metric. 
This implies that the differential map of $\cuSO{t}$ coincides with the 
Fermi--Walker transport along the worldline with respect to the \dS metric, 
cf.\ Definition~\ref{Fer} in the appendix. The desired solution is 
then obtained as follows. 

     Consider $(\dSm,g)$ as being embedded into ambient Minkowski space 
$(\Minm,\gm)$, {\it cf.\ }equation~\eqref{eqdSMin}, and denote 
$\curw{t}\in\dSm$ after this identification by $\cumin{t}$.
Then the acceleration 
of the observer in ambient Minkowski space is given by  
\begin{align} \label{eqam}
\am_t \doteq 
\frac{{\rm d^2}}{{\rm d}s^2}\, \cumin{t},\quad\text{ where $s=s(t)$ is the
  \dS proper time.} 
\end{align} 
Define now, for each $t\in\RR,$ a linear transformation of \Min{} by 
\begin{align} 
 M_t &\doteq  \gm(\dotcumin{t},\,\cdot\,)\,\am_t - 
\gm(\am_t,\,\cdot\,)\, \dotcumin{t} \,. \label{eqMt} 
\end{align}
Obviously, $M_t$ is skew-adjoint with respect to $\gm$. More specifically, 
in the Lorentz frame of $\dotcumin{t}\in\Minm$, $M_t$ 
is an infinitesimal boost with direction $\am_t$. With these definitions, 
we have: 

\begin{prop} \label{cuSO} 
Given a worldline $t\mapsto\curw{t}$ in \RW, there is precisely one curve 
$t\mapsto\cuSO{t}$ in $SO(4,1)$ satisfying equation~\eqref{eqgtx0}
and the above assumptions (a) and (b). It is given by 
\begin{equation}  \label{eqTexp}
 \cuSO{t} = T \exp \int_0^{t} M_s\,ds\,,  
\end{equation}
where $T$ denotes time ordering. 
\end{prop}

\paragraph{Remarks.} 1. Our assumptions, as well as the result, are
independent of the pa\-ra\-metri\-za\-tion of the worldline: If 
$\bar{\,\curw{t}}\doteq\curw{h(t)}$ is a reparametrization of the
worldline (where $h(0)=0$), then $\!\bar{\;\cuSO{t}}\doteq\cuSO{h(t)}$ 
is the corresponding solution in $SO(4,1)$. 
\\
2. One might impose, instead of (b), the weaker assumption that the 
measuring device  be irrotational only in the sense of
hydrodynamics~\cite{SaWu}. But in the present case of a rigid 
motion (described by de Sitter isometries $\cuSO{t}$) this weaker notion is  
equivalent to our assumption (b). 
\\
3. If $t\mapsto \curw{t}$ is a geodesic in \dS, then our curve 
$t\mapsto \cuSO{t}$ coincides with the well known  
corresponding  one-parameter group of boosts~\cite{NPT,BB}. 

\bigskip

     We prove the Proposition and the first remark in the appendix.

\section{Quantum Dynamics}

     We shall describe quantum systems in \RW by a net $\wnetrw$ of
$C^*$-algebras indexed by a suitable set of open subregions $\wRW$ of
\RW and a state $\orw$. The algebra $\Arw(\wrw)$ is interpreted as the
algebra generated by all observables measurable in the region $\wrw$,
and the state $\orw$ models the preparation of the quantum system.  We
emphasize that these data contain the same physical information
encoded in the standard formulation of quantum field theory in terms
of fields (cf. Section 4 of \cite{BH} and further references there).
The algebras are subalgebras of some $C^*$-algebra $\Arw$ on which the
state $\orw$ is defined. Since we shall only be concerned with a
single representation of this net, it will be no loss of generality to
consider the net $\wrnetrw$ with $\Rrw(\wrw) = \Arw(\wrw)''$ taken to be
the von Neumann algebra generated in the GNS representation space
$\Hs$ associated with $(\Arw,\orw)$. Hence, the unit vector 
$\Orw\in\Hs$ implementing the state $\orw$ on $\Arw$ is cyclic for $\Arw$.

     A {\em propagator family} is a family
$\{\proprw{t,s}\}_{t,s\in\RR}$ of automorphisms of the observable
algebra $\Rrw = \Arw''$ which satisfies the propagator identities
\begin{align} \label{eqProp} 
\proprw{t,s}\,\proprw{s,t'} =\proprw{t,t'}\quad\text{ and }\quad 
 \proprw{t,t} =\id \, , \quad\text{ for all } s,t,t'\in\RR 
\end{align} 
($\id$ denotes the identity map on $\Rrw$), and which is continuous in 
the following sense. Denoting by $\Bs(\Hs)$ the algebra of all bounded 
operators on the representation space $\Hs$, we require that the maps 
$\RR \ni t \mapsto \proprw{t,s}(A) \in \Bs(\Hs)$ be continuous for each 
fixed $s \in \RR$ and $A \in \Rrw$, with $\Bs(\Hs)$ endowed with the 
weak operator topology (and similarly with the roles of $t$ and $s$
interchanged).

     In order for an interpretation of this family as a dynamics for the 
observer to be physically reasonable, $t$ should be the observer's proper 
time and $\proprw{t,s}$ should map the observables measurable in the 
observer's immediate neighborhood at his proper time $s$ onto those 
measurable at time $t$. More precisely, let $t\mapsto \curw{t}$ be the 
worldline of our observer, and let, for each $t\in\RR,$ $\fadcrw{t}$ 
be a suitable neighborhood of $\curw{t}$ (interpreted as the space-time 
region wherein the measurement process initiated at time $t$ takes 
place. We shall say that 
the propagator family $\{\proprw{t,s}\}_{t,s\in\RR}$ is a 
{\em covariant dynamics} with respect to the family of space-time regions 
$\{\fadcrw{t}\}_{t\in\RR}$, if 
\begin{align} \label{eqpropCov}
 \proprw{t,s} \,\Rrw(\fadcrw{s}) = \Rrw(\fadcrw{t}) \, ,
\end{align}
for all $t,s\in\RR$. 

     It cannot be expected that the covariance property~\eqref{eqpropCov} 
will be strictly satisfied in all circumstances of physical interest.
There are two directions in which this property could be relaxed. First
of all, it could well happen that \eqref{eqpropCov} will only hold for
all $t,s$ in some finite open interval. In addition, the dynamics may
not maintain the localization of all of the observables localized in
$\fadcrw{t}$ but will do so for ``sufficiently many'' such observables.
For these reasons, if the propagator family satisfies
\begin{align} \label{eqpropQCov}
 \proprw{t,s} \,\Rrw_0(\fadcrw{s}) = \Rrw_0(\fadcrw{t}) \, ,
\end{align}
for all $t,s\in I$, for some open interval $I \subset \RR$, and 
for subalgebras $\Rrw_0(\fadcrw{t}) \subset \Rrw(\fadcrw{t})$ which are 
sufficiently large that the distinguished vector $\Orw$ is still cyclic for 
each of them, we shall say that the propagator family is a 
{\em quasi-covariant dynamics} with respect to  $\{\fadcrw{t}\}_{t\in I}$. 
Intuitively speaking, the forces described by such dynamics leave a 
large class of observables localized within the observer's laboratory, at 
least for some large but finite time interval but can act upon some of
the observables in such a way that their localization is moved out of 
the immediate neighborhood of the initial worldline, a situation which
one envisages in the presence of certain external fields.

\section{Examples Provided by Transplantation}

     In this section we provide specific examples fitting into the
above setting by employing the novel technique of transplantation of
theories from one space--time onto another \cite{BMS}. We consider a
fixed \RW space--time with embedding parameter $\alpha\leq \pi/2$, as
described in Section 2. For such a space--time we have described in
\cite{BMS} how one can obtain a net $\wrnetrw$ on \RW from a covariant
de Sitter theory via transplantation. We shall review that process
below for the reader's convenience. Given an arbitrarily accelerated
observer in \RW, we shall construct a propagator family acting on the
transplanted \RW theory and describing the time evolution of certain
distinguished observables in the observer's neighborhood, in the sense
of quasi-covariant dynamics. When $\alpha = \pi/2$, the constructed
dynamics is even covariant.

     It is necessary to summarize some of the results from our previous
paper \cite{BMS}. Consider the embedding~\eqref{eqdSMin} of \dS\, into 
Minkowski space and let $\Lid$ denote the proper orthochronous Lorentz group 
in five dimensions. Let $\wMin$ be the family of Minkowski space 
regions obtained by applying the elements of $\Lid$ to a single 
wedge--shaped region of the form 
\begin{equation}
\WRM \doteq \{ \xm \in \Minm \mid \xm_1 > \vert \xm_0 \vert \},
\end{equation}
\emph{i.e.\ }this family of regions is
$\wMin \doteq \{ \lambda \, \WRM \mid \lambda \in \Lid \}$.
Let then $\wdS$ be the collection of intersections
$\{ \wm \cap dS \mid \wm \in\wMin \}$. We call $\wdS$ the set of
de Sitter wedges. A wedge in de Sitter space is the causal
completion of the worldline of a freely falling observer. We define 
wedges in \RW to be the intersections with \RW of those de Sitter 
wedges whose edges\footnote{See below for a precise definition.} are 
contained in $RW$. They correspond to the causal completions of the union 
of worldlines of freely falling observers in $RW$. The set of these 
Robertson--Walker wedges will be denoted by $\wRW.$
\bigskip
\vspace*{-10mm}
\begin{figure}[ht] \label{F1}
\epsfxsize120mm
\epsfbox{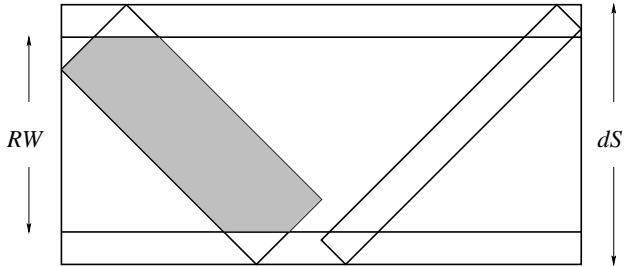}
\caption{Penrose diagram indicating a wedge in \dS (right) and in
  \RW (left).}
\end{figure}

We recall the following lemma \cite[Lemma 2.1]{BMS}.

\begin{lemma}\label{LemWRW} 
$\wRW$ is stable under the operation of taking causal
complements and under the action of the isometry group of
rotations $SO(4)$ on {\it RW}. Further, the elements of
$\wRW$ separate spacelike separated points in \RW and
$\wRW$ is a subbase for the topology in {\it RW}.
\end{lemma}

     One obtains a more intrinsic characterization of $\wRW$ by noticing  
that wedges in de Sitter space--time can be characterized by their edges.
Let $\ERM$ be the edge of $\WRM$, {\it i.e.\ }the three--dimensional 
subspace $\{\xm \in \Minm \mid \xm_0=\xm_1=0 \}$.
Applying the elements of $\Lid$ to $\ERM$, one obtains all 
three--dimensional spacelike linear subspaces $\EM$ of \Min. The
intersections of these with \dS are exactly the
two--dimensional, spacelike, totally geodesic, complete, connected
submanifolds of \dS. These submanifolds will be called
{\em  de Sitter edges} and are denoted by $E$. The causal complement
of $\ERM$ has two connected components, one being $\WRM$ and the
other one being its causal complement $\WRM{}'\in\wMin.$ 
Hence also the causal complement of any de Sitter edge
has two connected components, each being a wedge,
{\it i.e.\ }a Lorentz transform of $\WRM$ intersected with 
$dS$. So we conclude that the
wedges in \dS may be characterized as the connected components of the
causal complements of de Sitter edges.

     Based on this observation, we have given in \cite{BMS} an
analogous, more geometric characterization of wedges in a
Robert\-son--Wal\-ker space--time $RW$. 
 In fact, an {\it ad hoc} condition which was still left
in this characterization can be replaced by completely intrinsic
geometric constraints. Thus the choice we made above of wedges in \RW
is well-motivated from an intrinsic point of view. 
The details of this characterization will not be necessary in this
paper, however, and will be presented elsewhere. 

     In \cite{BMS} we defined a one--to--one map of the set $\wdS$ of 
de Sitter wedges onto the set $\wRW$ of Robertson--Walker wedges, which
we give again here. Recall 
that an element of the class of Robertson--Walker space--times
considered in this paper is embedded into \dS with a characteristic interval
$|\tau|< \alpha\leq\pihalf$.
If $\alpha =\pi/2$, then the embedded $RW$ coincides with $dS$.
In this case a Robertson--Walker wedge is just a de Sitter wedge and
the families $\wdS$ and $\wRW$ are identified by the
embedding. To cover the general case $0 < \alpha \leq \pi/2,$
we define a diffeomorphism $\Phi$ from
\dS onto \RW which bijectively maps the set of
de Sitter edges onto the set of Robertson--Walker edges. It is given by
\begin{equation} \label{5}
 \Phi \left(\tau,\chi,\theta,\phi\right) \doteq  (f(\tau),\chi,\theta,\phi),
\end{equation}
where\footnote{It is noteworthy that $f$ is uniquely 
fixed~\cite[Lem.\ A.2]{BMS} by the requirement that $\Phi$ be a
time-orientation preserving diffeomorphism from \dS onto \RW of the 
form~\eqref{5}, which bijectively maps the set of de Sitter edges
onto the set of Robertson--Walker edges.}
\begin{equation} \label{6}
f(\tau) \doteq
 \arcsin\big(\sin(\alpha)\,\sin(\tau)\big).
\end{equation}
The map $\Phi$ gives rise to a one--to--one correspondence
\begin{equation}  
\Xi: \wdS \rightarrow \wRW  
\end{equation}
as follows. Let $W$ be a de Sitter wedge with edge $E.$
The causal complement  $\Phi(E)'$ of $\Phi(E)$ in $RW$ has two connected
components, exactly one of which has nontrivial intersection with $W$.

\begin{definition}
Let $W$ be a de Sitter wedge with edge $E.$ We define $ \Xi(W)$
to be the connected component of $\Phi(E)'$ in $RW$ which has
nontrivial intersection with $W$.
\end{definition}

\noindent
The map $\Xi$ thus maps the family of de Sitter wedges onto
the family of Robertson--Walker wedges. It has the following
specific properties which we recall from \cite[Prop. 2.3]{BMS} and
which are crucial for our process of the transplantation 
of nets.

\begin{prop} \label{Prop3.1} 
The map $\Xi : \wdS \rightarrow \wRW$ is a
bijection.
It commutes with the operation of taking causal complements in the
respective spaces and
intertwines the action of the isometry group of rotations
SO(4) on $\wdS$ with its action on $\wRW$. 
\end{prop}

\noindent As shown in \cite{BMS}, if $\alpha<\pi/2,$
then $\Xi$ is not induced by a bijective point transformation from
\dS onto $RW,$ {\em i.e.\ }there is no map $p:\dSm\rightarrow\RWm$ such
that $\Xi(W) = \{ p(x) \mid x \in W \}$, for all $W\in\wdS$.

    We now consider an arbitrary covariant de Sitter theory having the
Reeh--Schlieder property. 
Such theories either can be constructed directly on de Sitter space or 
can themselves be transplanted from a covariant Minkowski space theory 
{\it via} a process akin to holography \cite{Fred,BMS}. Hence, we consider 
a net $\wrnetds$ of local algebras, a state vector $\Omega$ in the
representation space $\Hs$ and a strongly continuous unitary 
representation $U$ of the proper orthochronous de Sitter group $\Lid$ 
satisfying
\begin{equation}
U(\lambda)\Rs(W)U(\lambda)^{-1} = \Rs(\lambda W) \, , 
\end{equation}
as well as $U(\lambda)\Omega = \Omega$, for all $W \in \Ws$ and 
$\lambda \in \Lid$. The Reeh--Schlieder property refers to the 
condition that the vector $\Omega$ is cyclic for any local algebra 
$\Rs(\Os)$ corresponding to any (sufficiently small) 
bounded open region $\Os$ --- 
cf.\ \cite{BB}.\footnote{Such algebras either can already be provided 
in the net or can be constructed from the wedge algebras $\Rs(W)$ by suitable
intersections.} 

We will also assume that the de Sitter theory  satisfies the following 
physically motivated condition: Let $W \in \Ws$ be any de Sitter wedge
and $\alpha_t^{W}$, $t \in \RR$, the automorphism group induced on 
$\Rs(W)$ by the one-parameter subgroup of boosts which leaves $W$ 
invariant and induces a future directed Killing vector field in
$W$. Then $(\Rs(W),\alpha_t,\Omega)$ satisfies the KMS--condition
\cite{BratRob}, {\it i.e.\ }there exists some
$\beta > 0$ such that for any pair $A,B \in \Rs(W)$ there is an
analytic function $F$ in the strip 
$\{ z \in \CC \mid 0 < \text{Im}(z) < \beta\}$
with continuous boundary values at $\text{I}m(z) = 0$ and 
$\text{Im}(z) = \beta$, which are respectively given by
\begin{equation} \label{KMS}
F(t) = \langle\Omega, A\alpha_t(B)\Omega\rangle \; , \quad 
F(t + i\beta) = \langle\Omega, \alpha_t(B)A\Omega\rangle \; ,
\end{equation}
for all $t \in \RR$. As in \cite{BB}, we shall refer to this condition
as the {\it geodesic KMS--property} for brevity. The physical content of 
this assumption is that
to any uniformly accelerated observer in $W$ the de Sitter vacuum
must appear to be a thermal equilibrium state at inverse temperature
$\beta$. In fact, it has been shown in\cite{BB} that this temperature
{\it must} be the Gibbons--Hawking temperature.

     The geodesic KMS--property itself has
been proven to hold in many situations. For example, if (1) the de Sitter
theory satisfies the Condition of Geometric Modular Action and the
Modular Stability Condition \cite{BDFS,BMS}, or if (2) the de Sitter
theory is obtained from a five-dimensional Minkowski space theory
\cite{Fred,BMS} which itself is locally associated with a quantum
field satisfying the Wightman axioms \cite{BW}, then the geodesic
KMS--property is fulfilled \cite{BMS}. Or, as a third alternative, if the 
$n$-point functions of the de Sitter theory satisfy a certain
weak spectral condition \cite{BEM}, then the geodesic KMS--property is
satisfied, as long as the de Sitter quantum field is locally 
affiliated with the corresponding de Sitter algebras $\Rs(W)$.

     We proceed from the given net on $dS$ to a corresponding net
$\wrnetrw$ on \RW by {\it transplantation}, setting
\begin{equation} 
\Rrw(\wrw) \doteq \Rs(W),\quad \wrw = \Xi(W),
\end{equation} 
where $\Xi:\wdS\rightarrow \wRW$ is the bijection defined above. 
In addition, we proceed from $\Omega$ to a state vector $\Orw$ in the
\RW theory by defining
\begin{equation}
\Orw \doteq \Omega \, .
\end{equation}
We thus obtain from the original net and state vector describing a theory
on de Sitter space a net $\wrnetrw$ and state vector $\Orw$ describing
a theory on $RW$ which now are re-interpreted in terms of Robertson--Walker
space--time. The physical significance of the operators and state vector
thereby changes.

  {} For any (sufficiently small) bounded region  $\dcrw \subset \RWm$ we define
\begin{equation} \label{}
\Rrw(\dcrw) \doteq \bigcap_{\wrw\supset\dcrw} \,\Rrw(\wrw) \, .
\end{equation}
We remark that if the ``preimage'' in \dS of $\dcrw \subset \RWm$,
defined as 
\begin{equation} \label{eqXiDc}
 \Xi^{-1} (\dcrw) \doteq \bigcap_{\wrw\supset\dcrw} \,\Xi^{-1}(\wrw)  
\quad\subset \dS \,, 
\end{equation}
contains an open set, then $\Orw$ is cyclic for the algebra 
$\Rrw(\dcrw)$, since it contains $\Rs\big(\Xi^{-1}(\dcrw)\big)$. 
On the other hand, if $\alpha<\pi/2$ the set $\Xi^{-1}(\dcrw)$ may be empty 
if $\dcrw$ is too small~\cite{BMS}, see below. In this case, and if the
underlying de Sitter theory has the property that intersections of
algebras corresponding to disjoint wedges consist only of scalar
multiples of the identity\footnote{In \cite{TW} 
it is shown that this condition is satisfied under very general
circumstances.}, then $\Rrw(\dcrw)$ is trivial in the same sense.  
We shall consider, specifically, double cones; we
remind the reader that a double cone $\dcrw$ is a nonempty intersection of 
the interiors of a future lightcone and a past lightcone. A
particularly convenient class is comprised of the ``upright'' double cones 
in $RW$. Let $x=(\tau,\chi,\theta,\phi)\in\RWm$ and let
$\eps$ be a positive number such that $|\tau\pm\eps|<\alpha.$ 
To these data we associate a double cone 
\begin{equation} \label{eqDC}    
\dcrwsub{x,\eps} \doteq I_+(x_{-\eps})\cap I_-(x_{+\eps})\;,
\mbox{ where } x_{\pm\eps} \doteq (\tau\pm\eps,\chi,\theta,\phi)\;.     
\end{equation}
Here, $I_{\pm}(p)$ denotes the future, respectively 
past, light cone of a point $p$. By Proposition~2.5 of \cite{BMS}, 
$\Xi^{-1}(\dcrwsub{x,\eps})$ contains an open set if 
$\eps>\pi/2-\alpha$, but is empty if $\eps<\pi/2-\alpha$. 

     Given the curve $t\mapsto\cuSO{t}$ in $SO(4,1)$ associated to
the observer's worldline $t\mapsto \curw{t}$ by Proposition~\ref{cuSO}, 
we define
\begin{equation}  \label{eqprop1}  
\proprw{t,s} \doteq \Ad \,U(\cuSO{t}\cuSO{s}^{-1})
\quad\text{ for }s,t\in \RR\,.  
\end{equation} 
By the strong continuity and the representation property of $U$, this 
defines a propagator family. We shall see that it acts (quasi-)covariantly 
on certain observables in the observer's neighborhood. As it turns out, 
this propagator family provides a covariant dynamics 
in the special case where the embedding parameter $\alpha$ of the \RW
space--time under consideration equals $\pi/2$, a particular instance of
which is the de Sitter space--time itself. In fact, for geodesics in 
\dS this construction is well known \cite{NPT} and is generalized here 
to the case of worldlines corresponding to an arbitrarily accelerated 
motion in any \RW space--time conformally equivalent to $dS$. In contrast, 
if $\alpha<\pi/2$, the propagator family is a {\em quasi-covariant dynamics} 
with respect to  $\{\fadcrw{t}\}_{t\in I}$ and cannot be covariant. 

     We recall that when $\alpha = \pi/2$, the de Sitter group 
$SO(4,1)$ coincides with the group of conformal, orientation preserving 
transformations of $RW$, and the net of local algebras $\wrnetrw$
is covariant under the adjoint action of the representation 
$U$~\cite[Cor.\ 3.2]{BMS}. If $\dcrw$ is a neighborhood of $\curw{0}$, then 
\begin{equation} \label{eqOt1}
  \fadcrw{t}\doteq \cuSO{t}\dcrw 
\end{equation} 
is a neighborhood of $\curw{t}$, which we take as the space-time region 
representing the observer's laboratory at time $t$.  
The above-mentioned covariance then implies that $\proprw{t,s}$
satisfies equation~\eqref{eqpropCov} for this family of
neighborhoods, thus complying with the conditions for a covariant
dynamics. We therefore have proven the following.

\begin{prop} \label{Cov}
Let $t\mapsto \cuSO{t}$ be the curve in $SO(4,1)$ assigned to the
observer's worldline according to Proposition~\ref{cuSO}, and let
$\{\proprw{t,s}\}_{t,s\in\RR}$ be the corresponding propagator family
as defined in equation~\eqref{eqprop1}. If the embedding parameter 
$\alpha=\pi/2$, then it is a covariant dynamics with respect to every family 
of space-time regions of the form $\fadcrw{t}=\cuSO{t}\,\dcrw$, with 
$\dcrw$ a neighborhood of $\curw{0}.$ It describes the time evolution 
of observables measured in a non-rotating rest frame for the worldline. 
\end{prop}

     We now turn to the case $\alpha < \pi/2$. As indicated above, in this 
case there is a minimal length scale below which no nontrivial 
observables can be localized~\cite[Cor.\ 3.3]{BMS}, provided the de 
Sitter theory has the  trivial-intersection property mentioned above.  

In light of this fact, we require the space-time region $\fadcrw{t}$
representing the observer's laboratory at time $t$ (in the sense
indicated earlier) to contain the closure of the upright double cone 
$\dcrw_{\curw{t},\frac{\pi}{2}-\alpha}$ centered at $\curw{t}$ --- 
see equation~\eqref{eqDC}. We then say that $\fadcrw{t}$ is 
{\em sufficiently large for} $\curw{t}$, since $\Rrw(\fadcrw{t})$ is
large enough so that $\Orw$ is cyclic for it, 
{\it cf.\ }~\cite[Cor.\ 3.3]{BMS}. We shall further require that each 
$\fadcrw{t}$ is a double cone and that the family of double cones is 
continuous in
the sense that the two curves described by the future and
past apices of the elements of the family $\{\fadcrw{t}\}$ are both
continuous in $t$. If $\alpha=\pi/2$, the family of double cones
considered above (see equation~\eqref{eqOt1}) satisfies both of these
requirements. If $\alpha<\pi/2$, a continuous family of sufficiently large
double cones\footnote{If $\alpha < \frac{\pi}{2}$, such sufficiently large
double cones do not exist at all \cite[Cor. 3.3]{BMS}. As a
  matter of fact, in this case the smallest 
localization regions are wedges.}  can only
exist for $t$ in some finite interval $I$, since eventually the
corresponding future and past apices will reach the future,
respectively past, horizon of \RW.  Taking these basic facts into
account, we can establish the following result, whose proof is given
below.

\begin{theorem}  \label{Dyn} 
Let $t\mapsto \cuSO{t}$ be the curve in $SO(4,1)$ assigned to the
observer's worldline according to Proposition~\ref{cuSO}, and let
$\{\proprw{t,s}\}_{t,s\in\RR}$ be the corresponding propagator family
as defined in equation~\eqref{eqprop1}. If $\{\fadcrw{t}\}_{t\in   I}$,
with $I \subset \RR$ compact, is any continuous family of double cones 
such that $\fadcrw{t}$ is sufficiently large for $\curw{t}$, $t\in I$,
then $\{\proprw{t,s}\}_{t,s\in I}$ is a quasi-covariant dynamics for
this family.
\end{theorem}

     It is noteworthy that the subalgebras 
$\Rrw_0(\fadcrw{t})\subset \Rrw(\fadcrw{t})$ of observables, which
remain localized within the laboratory under the action of the
dynamics ({\it cf.\ }equation~\eqref{eqpropQCov}), can be
intrinsically characterized as follows. Given the 
time interval $I$ and the family of space-time regions $\fadcrw{t}$,
the desired algebras are given by 
\begin{equation} \label{eqWell} 
\Rrw_0(\fadcrw{t}) \doteq \bigcap_{s\in I}\; \proprw{t,s}\,
\Rrw(\fadcrw{s})\;,\quad t\in I\,.   
\end{equation}
This is a subalgebra of $\Rrw(\fadcrw{t})$, and for each 
$t,t'\in I$, the propagator $\proprw{t,t'}$ maps the algebra
$\Rrw_0(\fadcrw{t'})$ onto $\Rrw_0(\fadcrw{t})$ by virtue of the propagator 
identities~\eqref{eqProp}. What we claim is that these algebras are
sufficiently large such that the distinguished vector $\Orw$ is cyclic 
for each of them. 

\begin{proof} 
For fixed $t\in I$, consider the intersection of the 
$\tau$-coordinate line through 
$\curw{t}$ with the boundary of $\fadcrw{t}$. Since the latter consists 
of two achronal sets, this intersection consists of two points $x_\pm(t)$. 
Let $\eps_t$ be the minimum of the two numbers $|\tau(\curw{t})-\tau(x_+(t))|$ 
and $|\tau(\curw{t})-\tau(x_-(t))|$. Then, by hypothesis, one has
$\dcrw_{\curw{t},\eps_t}\subset \fadcrw{t}$ and 
$\eps_t>\frac{\pi}{2}-\alpha$. With $f$ the scaling function defined
in (\ref{6}), this implies that one has
$\hat{\eps}_t\doteq\pihalf-f^{-1}(\pihalf-\eps_t)>0$. Hence, Prop. 2.5
in \cite{BMS} entails that the double cone $\Os_{\curw{t},\hat{\eps}_t}$ 
is an open neighbourhood of $\curw{t}$ contained in $\Xi^{-1}(\fadcrw{t})$. 

     Now define the regions
\begin{equation} \label{eqOSec}
\Os_t\doteq \cuSO{t}^{-1}\,\Os_{\curw{t},\hat{\eps}_t} \quad\text{ and
  }\quad \Os\doteq \bigcap_{t\in I}\Os_t \,. 
\end{equation}
It will be shown that $\Os$ contains an open set. To this end denote by 
$y_t$ the point determined by the intersection of the 
future boundary of $\Os_t$ with the $\tau$-coordinate line containing 
$\curw{0}$. Since $\Os_t$ is an open neighbourhood of $\curw{0}$, 
$\tau(y_t)$ is strictly larger than $\tau(\curw{0})$, for each $t\in I$. 
Furthermore,  $t\mapsto \tau(y_t)$ is a continuous function due to 
the assumed continuity of the family $\{\fadcrw{t}\}_{t\in I}$.  
Hence by continuity of the function and compactness of the interval $I,$
the quantity 
\begin{equation} \label{eqDelta}
\delta\doteq \min_{t\in I} \{\tau(y_t)\} 
\end{equation}
is also strictly larger than $\tau(\curw{0})$. 
Now, by construction, all points on the $\tau$ coordinate line through 
$\curw{0}$ with $\tau$ coordinate in $[\tau(\curw{0}),\delta)$ are 
contained in $\Os.$  
Hence $\Os$ contains a timelike curve segment. Since $\Os$ is causally 
closed, this implies that $\Os$ contains an open set.

     By construction, $\Os$ satisfies 
$ \cuSO{t}\,\Os \subset \Xi^{-1}(\fadcrw{t})$, $t\in I$. Hence the local \dS 
algebra $\Rs(\cuSO{t}\Os)$ is contained in $\Rrw(\fadcrw{t})$, $t\in I$. 
Furthermore, the propagator $\proprw{t,s}$ maps $\Rs(\cuSO{s}\Os)$ onto 
$\Rs(\cuSO{t}\Os)$, $s,t\in I$, by covariance of the underlying \dS theory.  
These two facts imply that $\Rs(\cuSO{t}\Os)$ is contained in the 
algebra $\Rrw_0(\fadcrw{t})$, for each $t\in I$ (see equation~\eqref{eqWell}).
Since $\Os$ is sufficiently large,  
$\Orw=\Omega$ is cyclic for $\Rs(\cuSO{t}\Os)$, 
completing the proof that $\proprw{t,s}$ is almost covariant. 
\end{proof}

     To further bolster our claim that the propagator family is
describing a physically reasonable local dynamics, we recall 
the notion of a passive state of a $W^*$-dynamical system $(\Rs,\alpha_t)$. 
In \cite{PuWo} Pusz and Woronowicz give a definition\footnote{In point of
fact, Pusz and Woronowicz treat $C^*$-dynamical systems, but their 
work can be extended to $W^*$-dynamical systems in a manner which in
all essential respects is identical --- cf. \cite{BratRob}.} which is
motivated by an attempt to find an exact mathematical expression
for the physical idea that systems in equilibrium cannot perform
mechanical work in cyclic processes. We refer the reader to \cite{PuWo}
for that definition. For our purposes here, what is more useful
is the following condition, which Pusz and Woronowicz proved is
actually equivalent to the physically motivated definition. If
$\delta$ is the generator of the automorphism group 
$\{\alpha_t\}_{t \in \RR}$ representing the dynamics on $\Rs$,
{\it i.e.} if $\delta : \Rs \rightarrow \Rs$ is defined by
\begin{equation}
\delta(A) = \lim_{t \rightarrow 0} \frac{1}{t}(\alpha_t(A) - A) \, ,
\end{equation}
then a state $\langle\Orw, \cdot\,\Orw\rangle$ on $\Rs$ is passive 
(with respect to $\alpha_t$)  if and only if \cite{PuWo} 
\begin{equation} \label{eqpassivity}
-i \frac{d}{dt}\langle\Orw,U^* \alpha_t(U) \Orw\rangle \mid_{t=0} \; = \;
-i \langle\Orw, U^* \delta(U)\Orw\rangle \; \geq \; 0 \, ,
\end{equation}
for all $U$ in the connected component of the group of all unitary
elements of $\Rs$ in the uniform topology\footnote{Of course, $U$ must
also be in the domain of $\delta$.}. In the special case that 
$\alpha_t = \Ad\, e^{itH}$, with $H$ self-adjoint, equation (\ref{eqpassivity})
becomes
\begin{equation} \label{eqpassivity2}
\langle\Orw,U^* H U \Orw\rangle \; = \; \langle\Orw, U^* [H,U]\Orw\rangle \; 
= \; -i \langle\Orw, U^* \delta(U)\Orw\rangle \; \geq \; 0 \, ,
\end{equation}
where in the first equality we have used the fact that a passive
state is invariant under $\alpha_t$, for all $t \in \RR$ \cite{PuWo}.
     Pusz and Woronowicz have revealed the close connection between the 
KMS--condition and passivity \cite{PuWo}. And recent work has indicated
that passivity is closely related with other criteria for the stability of
quantum theories on curved space--times \cite{FV,SV}. See also
\cite{Kuckert1,Kuckert2}.

     Our dynamics is not expressed in terms of an automorphism group,
so we cannot expect the Second Law of Thermodynamics to have quite
the same translation into our setting as in that considered by
Pusz and Woronowicz. However, we find the following analogous
situation. For the specified worldline and the resultant curve
$\cuSO{t}$ in $SO(4,1)$ from Proposition~\ref{cuSO}, one has
for each fixed $s$
\begin{equation}
\proprw{t,s} = \Ad \,U(\cuSO{t}\cuSO{s}^{-1}) = 
\Ad \,U(\cuSO{t})U^{-1}(\cuSO{s}) \, ,
\end{equation}
hence the analogue of the left hand sides of (\ref{eqpassivity}) and 
(\ref{eqpassivity2}) is 
\begin{equation}  \label{eqproppassivity}
-i \frac{d}{dt} \, \langle\Orw,\, U^* \proprw{t,s}(U) \,\Orw\rangle
 \mid_{t=s} \; =  \langle\Orw, U^* H_s U \Orw\rangle \, ,
\end{equation}
where we have used Prop.~\ref{cuSO} and $H_s$ denotes the self-adjoint 
generator of the unitary one-parameter group $t\mapsto
U(\exp(tM_s))$. The quantity~\eqref{eqproppassivity} measures the energy 
transferred by the operation $U$ to the state in the local reference
frame of the observer. We claim that for each 
$s \in \RR$ this quantity is nonnegative for all unitaries $U$ in the algebra
$\Rrw(\Xi(W_s))$, where $W_s$ is the unique de Sitter wedge containing the 
point $x_s$ (viewed as an element of \dS after the embedding) and left 
invariant by the boost subgroup generated by $M_s$, {\it cf.\ }the
proof of Theorem~\ref{Passivity} below.  To be able to interpret this 
statement in terms of the Second Law, 
the operation $U$ should be performed in a neighborhood of $x_s$, 
{\it i.e.\ }the localization region $\Xi(W_s)$ of $U$ should contain 
$x_s$. This certainly holds at
the instant of time  when $x_s$ lies in the $\tau=0$ hypersurface 
(since then $x_s\in W_s$ implies $x_s\in \Xi(W_s)$) and, by
continuity, for some open time interval around this instance. 
If $\alpha = \pi/2$, then of course one has $\Xi(W_s) = W_s$, so that 
$x_s \in  \Xi(W_s)$ for {\em all} $s \in \RR$.

\begin{theorem}  \label{Passivity} 
Assume that the underlying de Sitter theory $(\wrnetds,\Omega)$ satisfies the
geodesic KMS--property, and let $\wrnetrw$ and $\Orw$ be obtained from
such a theory by transplantation, as above. 
Let $t\mapsto \cuSO{t}$ be the curve in $SO(4,1)$ assigned to 
the worldline $t\mapsto x_t$ by Proposition~\ref{cuSO}, and let
$\{\proprw{t,s}\}_{t,s\in\RR}$ be the corresponding propagator family
as defined in equation~\eqref{eqprop1}. 

    For each $s \in \RR$, the state induced by $\Orw$ on the net $\wrnetrw$ 
is quasi-passive for $\proprw{t,s}$ on the algebra $\Rrw(\wrw_s)$, in the
following sense:  The quantity in (\ref{eqproppassivity}) is nonnegative 
for all unitaries in $\Rrw(\wrw_s)$ which are in the domain of the 
indicated derivation,
where $\wrw_s = \Xi(W_s)$ and $W_s$ is the unique de Sitter wedge
on which the boost-subgroup generated by $M_s$ induces a future
directed flow. 
The wedge $W_s$ contains $x_s$ for all $s$ and $\wrw_s$ contains $x_s$ 
for $s$ in some open interval $I$ around the instant of time 
when the observer passes the $\tau=0$ hypersurface. If $\alpha=\pi/2$, 
then $I=\RR$. 
\end{theorem}

\begin{proof} 
\newcommand{\geod}{\gamma_s} 
Fix $s\in\RR$ and consider the one-parameter subgroup of $SO(4,1)$
generated by the boost $M_s$, {\it cf.\ }equation~\eqref{eqMt}. 
Denote by $W_s$ the unique wedge such that this subgroup
leaves $W_s$ invariant and induces a future directed timelike vector
field on $W_s$, and denote by $\alpha_t^{W_s}$, $t \in \RR$, the
corresponding automorphism group on $\Rs(W_s)$. 
{}From Theorem 1.2 of \cite{PuWo} the geodesic KMS--property implies
that the state induced on $\Rs(W_s)$ by $\Orw$ is 
passive with respect to the group $\alpha_t^{W_s}$, $t \in \RR$. Hence, 
equation (\ref{eqpassivity2}) and the $SO(4,1)$-invariance of $\Omega$ 
imply the non-negativity of the right hand side of equation 
(\ref{eqproppassivity}) for all unitaries in $\Rs(W_s)=\Rrw(\wrw_s)$ 
which are in the domain of the indicated derivation. 

\newcommand{\lm}{\tilde{l}} 
\newcommand{\dotxm}{\dot{\tilde{x}}} 
To show that $x_s\in W_s$,  it is convenient to consider the
ambient Minkowski space, denoting the 
point $x_s$ after the identification~\eqref{eqdSMin} by $\xm_s$ and
the Minkowski wedge corresponding to $W_s$ by $\wm_s$. 
Recall~\cite{BDFS} that every wedge $\wm$ in \Min{} can be
characterized by a pair of linearly independent lightlike vectors 
$\lm^+,\lm^-$ as the set of all $\xm$ satisfying $\gm(\xm,\lm^\pm)<0$. 
Consider now the two lightlike vectors 
$\lm_s^\pm\doteq |\am_s|^{-1} \am_s \pm |\dot{\xm}_s|^{-1}\dot{\xm}_s$, where 
$\am_s$ denotes 
the acceleration in ambient Minkowski space as in
equation~\eqref{eqam}, and $|\xm|\doteq |\gm(\xm,\xm)|^{1/2}$. 
By relation~\eqref{eqMt} one checks that  
$M_s\lm^\pm_s=\pm a\, \lm_s^\pm$ for $a=|\am_s||\dot{\xm}_s|>0$. 
Hence the boost subgroup generated by $M_s$ leaves the lightlike rays 
corresponding to $\lm_s^\pm$ invariant and generates a future-directed
flow on these rays. It follows that the pair $\lm_s^\pm$ characterizes
$\wm_s$ in the sense indicated above. Now
$\gm(\xm_s,\dot{\xm}_s)=0$ and $\gm(\xm_s,\am_s)=-1$, hence $\xm_s\in\wm_s$ as
claimed.  

As shown before the theorem, also $\wrw_s = \Xi(W_s)$ contains $x_s$ for
$s$ in some interval $I$.  Note however that, if $\alpha\neq \pi/2$, 
this interval is in general not the entire real line, not even if the
worldline is a \RW geodesic.
\end{proof}

\section{Conclusion and Outlook}

     Given the worldline of any observer in \RW, we have constructed a
curve in $SO(4,1)$ which generates the worldline of the observer and
of neighboring worldlines. Our curve is uniquely fixed by the
requirement that the latter worldlines describe, potentially, material
particles moving along with the observer without colliding with him or
rotating with respect to him. With the same assumptions,
we show that this curve determines a transport of vectors along the 
worldline which coincides with the Fermi--Walker transport\footnote{This
last point is proven at the end of the appendix.}. 

     We proposed to implement the dynamics associated with this
transport using propagator families, and then we produced a large
class of examples of nets and states admitting such dynamics. We also
have shown why it is necessary to allow for dynamics of the type we
call quasi-covariant. In these examples we have also demonstrated that
the states are locally quasi-passive with respect to the  constructed dynamics,
so these states behave locally as if they were equilibrium states with
respect to the quantum dynamics we have associated with {\it arbitrarily
accelerated} observers on the class of Robertson--Walker space--times
which can be conformally embedded into de Sitter space.

     We anticipate that this process of transplantation of theories from
one space--time to another can also be performed for those 
Robertson--Walker space--times which can be conformally embedded into
Minkowski space, as well as for all the space--times which can be
conformally embedded into Einstein space. We then expect that 
quasi-covariant dynamics can be constructed for such theories along
the lines followed in this paper. 

     In \cite{Keyl1,Keyl2} Keyl takes another approach to finding
dynamics for quantum systems on curved space--times. Though his
starting point is also the worldline of an observer, the ``dynamical
maps'' he constructs are *-isomorphisms between pairs of local
algebras  associated with double
cones centered on the worldline at different times and having the same
extension in the time direction. They therefore must satisfy
propagator identities which differ from ours. He only considers the
analogue of what we call covariant dynamics. Although Keyl's approach
has the advantage over (the present state of) our proposal of being
applicable to any globally hyperbolic space--time, his construction of
dynamics uses in an essential manner properties of free field theory. 

     We close with a final comment. It is well known that in the
standard passage from classical Lagrangian field theory to quantized
field theory, symmetries of the classical Lagrangian can be broken in
the quantized theory. What is less well known is that a quantum field
theory can have symmetries which have no counterpart in the classical
theory --- modular symmetries associated with the local algebras
$\Rs(\Os)$ and a state vector $\Omega$ by the Tomita--Takesaki modular
theory (cf. \cite{Tak,BratRob1}). These modular symmetries may be 
implemented by the modular unitaries or, more generally\cite{BDFS}, by
the modular conjugations.  The consequences and uses of
these symmetries are not yet well understood, though some indications
are beginning to emerge \cite{Bor,SchrWies,BDFS,FS}. As was shown in
\cite{BMS}, the representation $U(\lambda)$ of $SO(4,1)$ used in this
paper to construct the quasi-dynamics on the Robertson--Walker net of
observable algebras consists of modular symmetries of the \RW theory
constructed above and can therefore be derived from the \RW theory
without any reference to the \dS theory. Hence, we have demonstrated
in this paper that such modular symmetries may be used to define a
dynamics for a quantum system, even though they have no relation to
the isometries of the underlying space--time, indeed though they are
not even implemented by point transformations on the space--time.

\appendix

\section{Proof of Proposition~\ref{cuSO}}

     In the following discussion  we shall have
recourse to the Fermi--Walker transport, which formalizes in classical
relativity theory the notion of irrotational transport along an
observer's worldline. For the reader's convenience, we recall its
definition. We will denote the tangent space to a manifold $M$ at
$x\in M$ by $T_xM$ and the differential map of a map
$\phi:M\rightarrow N$ at $x\in M$ by $T_x\phi$.

\begin{definition} \label{Fer} 
Let $t\mapsto x_{t}$ be a worldline in a space-time $(M,g)$. 
The Fermi--Walker derivative of a vector field $V$ along the worldline 
$t\mapsto x_{t}$ is defined by 
\begin{align} \label{eqFer}
F_{\dot{x}_{t}}V \doteq \nabla_{\dot{x}_{t}} V + g(a_t,V) 
\dot{x}_{t} -g(\dot{x}_{t},V)a_t\,,  
\end{align} 
where $a_t$ is the corresponding acceleration: 
$a_t\doteq 
\nabla_{x'_{t}}x'_{t}$, with $x'_t$ denoting the normalized tangent
vector, $x'_t\doteq g(\dot{x}_{t},\dot{x}_{t})^{-1/2}\, \dot{x}_t$. 
 
     A vector $v_t\in T_{x_t}M$ is said to arise from $v_0\in T_{x_0}M$ by 
Fermi--Walker transport along the worldline if there is a vector field
$V$ on the worldline satisfying $V_{x_0}=v_0,$ $V_{x_t}=v_t$ 
and $F_{\dot{x}_{s}}V =0$ for all $s\in [0,t].$ 
\end{definition}

     It follows from the definition that the Fermi--Walker transport is an 
isometry and preserves the normalized tangent vector to the curve 
(in the sense of equation~\eqref{eqLamdotx} below). Further, the
linearity property 
\begin{equation} \label{eqLin}
 F_{f(t)\,\dot{x}_{t}}V=f(t)\,F_{\dot{x}_{t}}V 
\end{equation}
implies that the Fermi--Walker transport is independent of the
parametrization. That is to say, if $\bar{x}_{t}$ is 
a reparametrization of the worldline, then the Fermi--Walker transports from 
$x_{0}$ to $x_{t_0}$ along the curves $t\mapsto \bar{x}_{t}$ and 
$t\mapsto x_{t}$ coincide.  

{}For the proof of Proposition~\ref{cuSO} we  proceed in two steps. 
First, we show that any differentiable curve 
$t\mapsto \cuSO{t}$ satisfying equation~\eqref{eqgtx0} and 
conditions (a) and (b) in Section 2 has to satisfy the initial
condition $\cuSO{0}=1$ and the differential equation 
\begin{align}  \label{eqldot} 
\frac{{\rm d}}{{\rm d}t}\cuSO{t} &= M_t  \circ \cuSO{t}, \quad
\text{ for all } t\in\RR\,, 
\end{align}
where $M_t$ is given by \eqref{eqMt}. It therefore is unique. Then we
verify that the solution \eqref{eqTexp} of this equation indeed
complies with the above constraints. 

     Thus, let $t\mapsto \cuSO{t}$ be a differentiable curve 
in $SO(4,1)$ satisfying equation~\eqref{eqgtx0} and assumptions (a) 
and (b). Since these assumptions
involve only the conformal, not the metric, structure, 
$t\mapsto\curw{t}$ may be considered as a curve in \dS. As is well known
and will be shown subsequent to this proof for the convenience of the
reader, the constraints (a) and (b) are equivalent to the condition 
that the differential map $T_{\curw{0}}\cuSO{t}$ of $\cuSO{t}$  
coincides with the Fermi--Walker transport, with respect to the \dS metric, 
from $\curw{0}$ to $\curw{t}$ along the worldline. Hence, one has
\begin{align} \label{eqFerDif} 
 F_{\dotcurw{t}}\big(T_{\curw{0}}\cuSO{t}\cdot v\big)&=0 
\quad \text{ for all } v\in T_{\curw{0}}\dS,t\in\RR\,,\\
T_{\curw{0}}\cuSO{0}&=1  \quad\text{ (initial condition)}\,.\label{eqIni}
\end{align} 

     To evaluate these equations, we consider \dS as being embedded into 
ambient $5$-dimensional Minkowski space \Min, 
{\it cf.\ }equation~\eqref{eqdSMin}, and denote $x\in\dSm$ after this  
identification by $\xm$.
Then the tangent space at a point $\xm\in \dS$ can be identified with 
the orthogonal complement $\xm^\perp$ of $\xm$. 
Further, the action of $SO(4,1)$ on 
\dS is just the restriction of its linear action on \Min, and the 
differential map $T_{\curw{0}}\cuSO{t}$ is just the linear map 
$\cuSO{t} : \cumin{0}^{\perp} \rightarrow (\cuSO{t}\cumin{0})^\perp$. 
Denote by $\gm$ and $\nablam$ the metric and covariant derivative in \Min,
respectively. For vector fields $X,Y$ on \dS one has\cite[Lemma 4.27]{O}
\begin{equation} \label{eqNabNabm}
 \nablam_XY=\nabla_XY+\gm(X,Y)\,P \,,
\end{equation}
where $P$ denotes the position vector field $P_\xm\doteq \xm\in\Minm$. 
Using this equation, one readily verifies that the (restriction of
the) Fermi--Walker derivative of ambient \Min{} coincides with that
of  \dS. Hence equation~\eqref{eqFerDif} is equivalent to 
\newcommand{\vm}{{\tilde{v}}}
\begin{equation}  \label{eqFer'}
\nablam_{\dotcumin{t}} (\cuSO{t} \vm) =  \gm(\dotcumin{t},\cuSO{t}
\vm)\,\am_t - \gm(\am_t,\cuSO{t} \vm)\, \dotcumin{t} \quad\text{
   for all } \vm\in \cumin{0}^\perp\,,\;t\in\RR\,,  
\end{equation}
where $\am_t\doteq \nablam_{\cumin{t}'}\cumin{t}'= 
\frac{{\rm d^2}}{{\rm d}s^2}\,
\cumin{t}
$ as in equation~\eqref{eqam}. On the other hand, one has 
\begin{equation}  \label{eqDer}
 \nablam_{\dotcumin{t}} (\cuSO{t}\vm) = \frac{{\rm d}}{{\rm d}t}
(\cuSO{t}\vm) = (\frac{{\rm d}}{{\rm d}t}\cuSO{t})\vm\,. 
\end{equation}
Since $\vm\in \cumin{0}^\perp$ was arbitrary and $\cuSO{t}$ is fixed by
its action on $\cumin{0}^\perp$, equation~\eqref{eqFer'} can therefore
be written as the differential equation~\eqref{eqldot} for $\cuSO{t}$.  

 The initial condition~\eqref{eqIni}, translated to ambient \Min, reads 
$\cuSO{0}=1$ on $\cumin{0}^\perp.$  
Hence, in view of $\cuSO{0}\cumin{0}=\cumin{0},$ we find that $\cuSO{0}=1$.
As $s\mapsto M_s$ is continuous, the unique solution to the differential 
equation~\eqref{eqldot}, with initial condition $\cuSO{0}=1$, is
given by equation~\eqref{eqTexp}. Moreover, since $M_t$ is  skew-adjoint with
respect to $\gm$ for each $t$, it lies in $SO(4,1)$.
It remains to show that this solution indeed satisfies 
$\cuSO{t} \cumin{0}=\cumin{t}$ for all $t\in \RR$ , as required by 
equation~\eqref{eqgtx0}. To this end, let 
\newcommand{\ym}{{\tilde{y}}}    
$\ym\in\Minm.$  Then one has
\begin{equation} \label{eqddt}
\frac{\rm d}{{\rm d}t}\gm(\cumin{t},\cuSO{t}\ym)
=\gm(\dotcumin{t},\cuSO{t}\ym)+\gm(\cumin{t},M_t\cuSO{t}\ym)
=\gm(\dotcumin{t},\cuSO{t}\ym)\big(
1 +\gm(\cumin{t},\am_t)\big)\,, 
\end{equation}
where relation~\eqref{eqMt} and  $\gm(\cumin{t},\dotcumin{t})=0$ was
used in the second equation. 
By equation~\eqref{eqNabNabm}, the vector $\am_t-\cumin{t}$ is in
$T_{\curw{t}}dS=\cumin{t}^\perp,$ hence one sees that
$\gm(\cumin{t},\am_t)=\gm(\cumin{t},\cumin{t})=-1.$  The right hand side of
equation~\eqref{eqddt} is therefore zero, which implies that 
$\gm(\cumin{t},\cuSO{t}\ym)=\gm(\cumin{0},\ym)$. It follows, in particular,
that $\cuSO{t}$ maps $\cumin{0}^\perp$ to $\cumin{t}^\perp$. 
Since $\cuSO{t}$ is a time-orientation preserving isometry, this
implies that $\cuSO{t} \cumin{0}= \cumin{t}$, completing the proof of
the Proposition. 

     We next prove our assertion that assumptions (a) and (b) are 
equivalent to the condition that $T_{\curw{0}}\,\cuSO{t}$ coincides with 
the Fermi--Walker transport, which we denote by $\Ts_{t,0}$, $t\in\RR$.  
The first step is to show that assumption (a) implies that the differential 
map preserves the normalized (with respect to the \dS metric) tangent 
vector to the world line, as does the Fermi--Walker transport : 
\begin{equation} 
 T_{\curw{0}}\cuSO{t} \cdot g(\dotcurw{0},\dotcurw{0})^{-1}\,
\dotcurw{0} \, = \,  g(\dotcurw{t},\dotcurw{t})^{-1} \,
\dotcurw{t}\,.  \label{eqLamdotx}
\end{equation}
Note that $\cuSO{t} \curw{s}$ must lie on the worldline of the
observer for all $t\in\RR$ and sufficiently small $s$. 
(Otherwise, $t\mapsto\cuSO{t} \curw{s}$ would be a neighboring 
worldline colliding with the observer at $\curw{s}$, contradicting 
assumption (a).) That is to say, $\cuSO{t} \curw{s}=\curw{f(t,s)}$ for 
some function $f$. Hence, $T_{\curw{0}}\cuSO{t}\cdot\dotcurw{0}$ is 
of the form $\partial_2f(t,0)\;\dotcurw{t}$, and is, in particular, 
collinear with $\dotcurw{t}$. Since $\cuSO{t}$ preserves the metric and 
time-orientation, this implies equation~\eqref{eqLamdotx}. 

     Suppose now that the axis of a gyroscope carried by the observer 
points towards some neighboring worldline at time $t=0$. 
The corresponding spatial direction is a 
unit vector in $\dotcurw{0}^\perp$, which we denote by $e.$ As is 
well known, the motion of the gyroscope axis corresponds to the 
Fermi--Walker transport along the worldline~\cite{MTW,SaWu}, 
{\it i.e.\ }its direction at time $t$ then corresponds to 
$\Ts_{t,0}\cdot e\in \dotcurw{t}^\perp.$ On the other hand, the 
direction of our measurement device at time $t$ corresponds, by assumption 
(a), to the projection of $T_{\curw{0}}\cuSO{t}\cdot e$ onto
$\dotcurw{t}^\perp$~\cite{SaWu}. But as a consequence of 
equation~\eqref{eqLamdotx}, the latter coincides with 
$T_{\curw{0}}\cuSO{t}\cdot e$.  Hence, assumption (b) implies that 
$T_{\curw{0}}\,\cuSO{t}=\Ts_{t,0}$ on $\dotcurw{0}^\perp.$
Since the Fermi--Walker transport also acts on $\dotcurw{0}$ like
$T_{\curw{0}}\cuSO{t}$ (see equation~\eqref{eqLamdotx}), we
have shown that (a) and (b) imply 
$T_{\curw{0}}\,\cuSO{t}=\Ts_{t,0}$, $t\in\RR$. 
The converse implication is also clear from our discussion. 

     We now show that the result is independent of the parametrization, as
remarked after the Proposition. To this end, let 
$\barcurw{t}\doteq\curw{h(t)}$ be a reparametrization of the
worldline (with $h(0)=0$), and $\barcuSO{t}$ the corresponding 
solution of $\barcuSO{t} \,\barcurw{0}=\barcurw{t}$
complying with assumptions (a) and (b). Since the Fermi--Walker
transport is independent of the parametrization, (a) and (b) still 
imply that the differential map of $\barcuSO{t}$ 
coincides with the Fermi--Walker transport, with respect to the
\dS metric, from $\barcurw{0}$ to $\barcurw{t}$ along
$t\mapsto\barcurw{t}$. Hence, equations~\eqref{eqFerDif} and
\eqref{eqIni} must hold, with $\curw{t}$ and $\cuSO{t}$ replaced by
$\barcurw{t}$ and $\barcuSO{t}$. Since both the Fermi--Walker
derivative and the covariant derivative are function-linear as in
equation~\eqref{eqLin}, the line of arguments leading to
equation~\eqref{eqldot} now leads to 
\begin{equation} \label{eqbarldot}
\frac{{\rm d}}{{\rm d}t}\barcuSO{t} = h'(t){M}_{h(t)}  \circ \barcuSO{t}\quad
\text{ for all } t\in\RR\,,
\end{equation}
with $M_t$ as in equation~\eqref{eqMt}. The unique solution with the
correct initial condition $\barcuSO{0}=1$ is given by 
$\barcuSO{t} = T \exp \int_0^{h(t)} M_s\,ds=\cuSO{h(t)}$, as
claimed.

\bigskip

\noindent {\bf Acknowledgements}:  DB and SJS wish to thank the Department
of Mathematics of the University of Florida, Gainesville, and the 
Institute for Theoretical Physics at the University of G\"ottingen,
respectively, for hospitality and financial support. DB and JM are
also grateful for financial support by the Deutsche Forschungsgemeinschaft 
(DFG). 
\\

\end{document}